\documentclass[letter]{aa} 

\usepackage{graphicx}
\usepackage{txfonts}
\newcommand{\um}{\,$\mu$m}
\newcommand{\kms}{km\,s$^{-1}$}
\newcommand{\cc}{cm$^{-3}$}
\newcommand{\oi}{[O\,{\sc i}]}
\newcommand{\ci}{[C\,{\sc i}]}
\newcommand{\cii}{[C\,{\sc ii}]}
\newcommand{\twcii}{[$^{12}$C\,{\sc ii}]}
\newcommand{\thcii}{[$^{13}$C\,{\sc ii}]}

\newcommand{\thco}{$^{13}$CO}
\newcommand{\twc}{$^{12}$C}
\newcommand{\thc}{$^{13}$C}
\newcommand{\cplus}{C$^+$}
\newcommand{\twcplus}{$^{12}$C$^+$}
\newcommand{\thcplus}{$^{13}$C$^+$}
\begin{document}

   \title{First detection of \thcii\ in the Large Magellanic Cloud}

   \author{Yoko Okada \inst{1}
           \and
           Ronan Higgins \inst{1}
           \and
           Volker Ossenkopf-Okada \inst{1}
           \and
           Cristian Guevara \inst{1}
           \and
           J\"{u}rgen Stutzki \inst{1}
           \and
           Marc Mertens \inst{1}
   }

   \institute{I. Physikalisches Institut der Universit\"{a}t zu K\"{o}ln, Z\"{u}lpicher Stra{\ss}e 77, 50937 K\"{o}ln, Germany
              \email{okada@ph1.uni-koeln.de}
   }

   \date{}

 
  \abstract
   {\thcii\ observations in several Galactic sources show that the fine-structure \twcii\ emission is often optically thick (the optical depths around 1 to a few).}
   {We want to test whether this also affects the \twcii{} emission from nearby galaxies like the Large Magellanic Cloud (LMC).}
   {We observed three star-forming regions in the LMC with upGREAT on board SOFIA at the frequency of the \cii\ line. The 4GHz band width covers all three hyperfine lines of \thcii\ simultaneously. For the analysis, we combined the \thcii\ F=1-0 and F=1-1 hyperfine components, as they do not overlap with the \twcii\ line in velocity.}
   {Three positions in N159 and N160 show an enhancement of \thcii\ compared to the abundance-ratio-scaled \twcii\ profile. This is likely due to the \twcii\ line being optically thick, supported by the fact that the \thcii\ line profile is narrower than \twcii, the enhancement varies with velocity, and the peak velocity of \thcii\ matches the \oi\ 63\um\ self-absorption. The \twcii\ line profile is broader than expected from a simple optical depth broadening of the \thcii\ line, supporting the scenario of several PDR components in one beam having varying \twcii\ optical depths. The derived \twcii\ optical depth at three positions (beam size of 14\arcsec, corresponding to 3.4~pc) is 1--3, which is similar to values observed in several Galactic sources shown in previous studies. If this also applies to distant galaxies, the \cii\ intensity will be underestimated by a factor of approximately 2.}
   {}


   \maketitle
%

\section{Introduction}

The \cii\ 158\um\ line ($^2$P$_{3/2}$-$^2$P$_{1/2}$ fine structure line) is one of the dominant cooling lines in photon-dominated regions \citep[PDR,][]{TH85I,Sternberg1995,Roellig2006}. Because of its close link to the UV radiation from massive stars, the \cii\ emission is often used to trace star formation in distant galaxies. PDR models show that the \cii\ emission typically stems from a depth of up to $A_V\sim 1$ on the PDR surface, which corresponds to a \cii\ optical depth of around unity\footnote{e.g. the peak optical depth is 0.8 using an extinction cross section from \citet{Draine2003a} for $R_V=4$, a carbon abundance of $1.6\times 10^{-4}$ \citep{Sofia2004}, an excitation temperature of 50~K and the line width of 1.5\kms.} \citep{Ossenkopf2013}. In order to quantify the optical depth of \twcii, we need to compare it with an optically thin line. This is naturally given by the \thcii\ lines. Because the \thcii\ $^2$P$_{3/2}$-$^2$P$_{1/2}$ transition splits into three hyperfine components and they are within $\pm 65$~\kms\ of the \twcii\ line \citep{Cooksy1986}, high resolution spectroscopy with a good sensitivity is needed to detect \thcii. After detections of the \thcii\ hyperfine components in M42 \citep{Boreiko1988,Stacey1991,BoreikoBetz1996b}, the Heterodyne Instrument for the Far-Infrared (HIFI) on {\it Herschel} and the German REceiver for Astronomy at Terahertz Frequencies (GREAT\footnote{GREAT is a development by the MPI f\"{u}r Radioastronomie and the KOSMA / Universit\"{a}t zu K\"{o}ln, in cooperation with the MPI f\"{u}r Sonnensystemforschung and the DLR Institut f\"{u}r Planetenforschung}) on board the Stratospheric Observatory for Infrared Astronomy (SOFIA) enable us to detect the \thcii\ lines in more Galactic sources \citep{Graf2012,Ossenkopf2013,Goicoechea2015,Guevara2019}. \citet{Guevara2019} showed that a uniform excitation temperature model gives an optical depth of 2 to 7 in four Galactic sources, as a lower limit of the optical depth due to the limited signal-to-noise ratio (S/N) of the \thcii\ spectra and the high excitation temperature assumption. A multi-layer model including an absorption layer requires higher optical depths.

In this letter, we report the first detection of \thcii\ emission from the Large Magellanic Cloud (LMC). 

\section{Observation and data reduction}

\begin{table*}
\caption{Summary of the pointed observations for the \thcii\ line.}
\label{table:obssummary}
\centering
\begin{tabular}{cccccccc}
\hline\hline
Position ID & Region & pixels & \multicolumn{2}{c}{position (J2000)} & $t_{\textrm{ON}}^\textrm{a}$ [min] & $\sigma_{\textrm{rms}}^\textrm{b}$ [K] & $T_\textrm{mb,12}/T_\textrm{mb,13,tot}^\textrm{c}$ \\
\hline
1 & N159 & LFAH\&V\_PX00 & 05:39:37.8 & -69:46:09.9 & 28.9 & 0.06 &  23\\
2 & N159 & LFAH\&V\_PX05 & 05:39:36.9 & -69:45:35.8 & 28.9 & 0.07 & (16)$^\textrm{d}$\\
3 & 30Dor & LFAH\&V\_PX00 & 05:38:48.8 & -69:04:43.0 & 10.8 & 0.11 & (15)$^\textrm{e}$\\
4 & N160 & LFAH\&V\_PX00 & 05:39:39.2 & -69:39:06.7 & 59.5 & 0.04 & 30\\
5 & N160 & LFAH\&V\_PX04 & 05:39:43.8 & -69:38:41.6 & 59.5 & 0.06 & 17\\
\hline
\end{tabular}
\begin{list}{}{\setlength{\itemsep}{0ex}}
\item[$^\textrm{a}$]The on-source integration time, same as the OFF integration time.
\item[$^\textrm{b}$]The velocity range used to derive $\sigma_{\textrm{rms}}$ is 130--145~\kms, 195--210~\kms, 260--273~\kms, and 323--338~\kms\ for N159, 125--145~\kms\ and 323--343\kms\ for 30Dor, and 115--135~\kms\ and 333--353~\kms\ for N160.
\item[$^\textrm{c}$]At the velocity of the \thcii\ peak.
\item[$^\textrm{d}$]$T_\textrm{mb,13,tot}$ peak is less than $3\sigma_{\mathrm{rms}}$ of the combined spectra.
\item[$^\textrm{e}$]Possibly affected by the wing of the \twcii\ line.
\end{list}
\end{table*}

\begin{figure*}
\centering
\includegraphics[width=0.9\hsize]{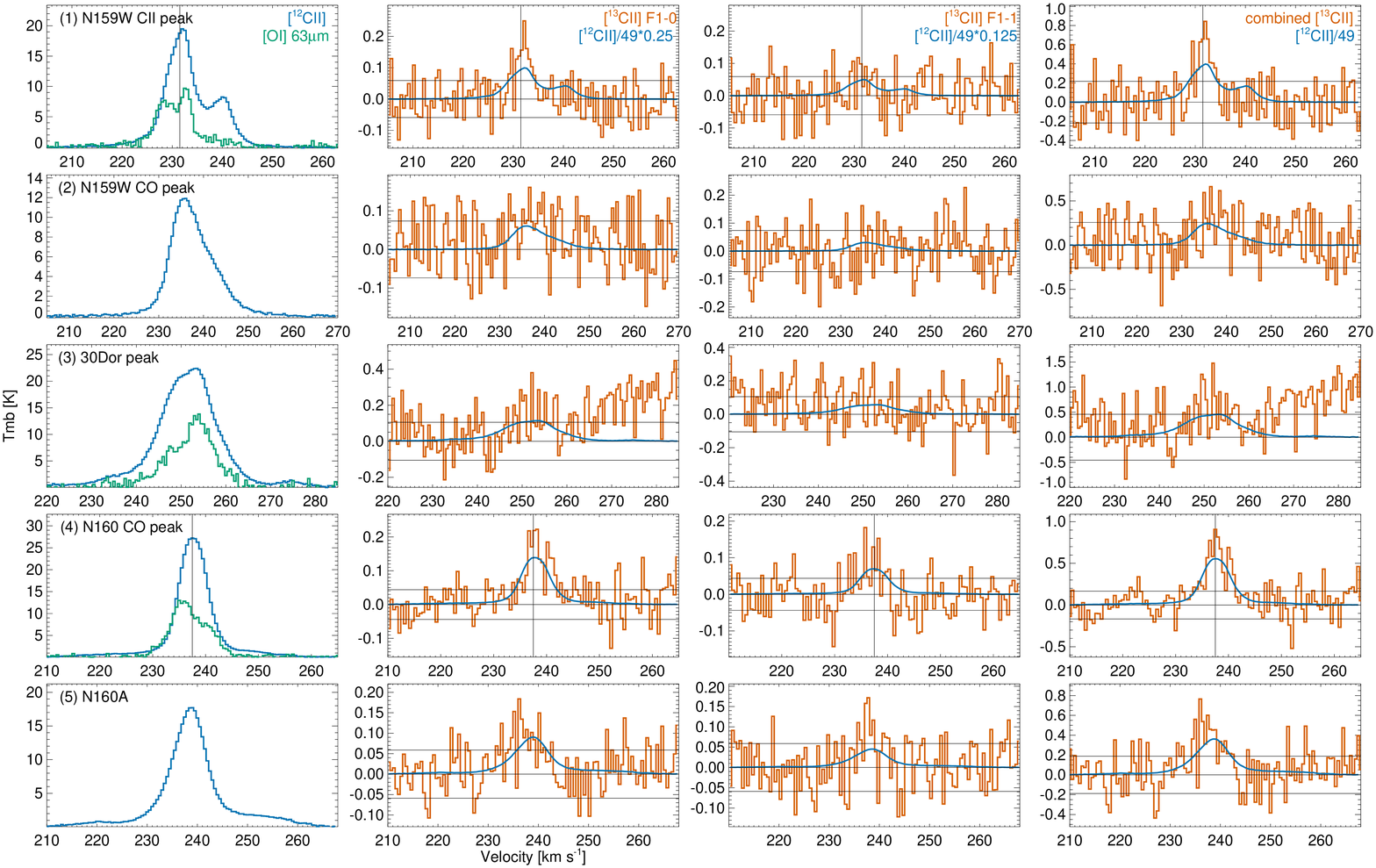}
\caption{Spectra of \thcii, \twcii\ and \oi\ at selected positions in the LMC. (Left) \twcii\ spectra (blue) and \oi\ 63\um\ spectra when available (green). The vertical lines guide to compare the velocities of the line profiles. (Middle two panels) \thcii\ F=1-0 and F=1-1 spectra (red) and \cii\ spectra (blue) scaled for optically thin emission and \twcplus/\thcplus$=49$. The horizontal lines indicate the rms noise of the baseline. (Right) Combined \thcii\ spectra (red) and the scaled \cii\ spectra (blue). See text for the formula of the combined \thcii\ spectra and the scaled \cii\ spectra.}
\label{figure:spectra_selected}
\end{figure*}

We observed \twcii\ and the \thcii\ hyperfine lines at selected positions in N159, 30~Dor, and N160 with upGREAT \citep{Risacher2016,Heyminck2012} on board SOFIA \citep{Young2012} in June 2018, as part of the guaranteed time in cycle 6 observations during two flights. The two polarizations of the Low Frequency Array (LFAH and LFAV) were tuned to the \cii\ line at 1900.5369~GHz. The band width of 4~GHz allows us to observe all three \thcii\ hyperfine emission lines simultaneously. In all positions, the strongest \thcii\ line (F=2-1) is blended with the \twcii\ line, so that we ignored this line in the analysis. In parallel, the High Frequency Array (HFAV) was tuned to the \oi\ 63\um\ line. Each array has seven pixels in a hexagonal configuration. The beam size is 14\arcsec\ for \cii\ and 6.3\arcsec\ for \oi, and the pixel separation in the hexagonal configuration is scaled by the beam size in the two frequency arrays, sharing the position of the center pixels. The central pixels of LFAH and LFAV are aligned within $\sim 2$\arcsec, and the central pixel of HFAV is about 3\arcsec\ away from the LFA central pixels. In the first flight, we started to observe N160 with single-phase chopped observations, and switched to the double beam switch mode later to have a better baseline in some of the pixels. In the second flight, we observed N159, together with a short observation of 30~Dor in double beam switch mode. The position in 30~Dor had a lower priority because its \twcii\ line profile is very broad and it overlaps with the \thcii\ F=1-0 line (see Fig.~\ref{figure:spectra_selected}; the 30~Dor spectrum at $>270$~\kms\ in \thcii\ F=1-0 shows a wing of \twcii.)

The data were calibrated by the standard GREAT pipeline \citep{Guan2012}, which converts the observed counts to the main beam temperature scale ($T_{\mathrm{mb}}$). For the N160 data, we confirmed that the baseline structures in the single-phase observations match those in the spectra of the same phase (phase A) extracted from the double beam switch observations. Therefore we averaged the spectra of the other phase (phase B) from the double beam switch observations, and added them to the single-phase observations to obtain better baselines in a few pixels. Although the S/N of the result is limited by the integration time of the phase B observations, we still gain S/N compared to ignoring the single phase data. We subtracted linear baselines and spectrally resampled the data to 0.5~\kms\ channel width. We then averaged the spectra at each position and each pixel weighting them by the baseline noise. In Table~\ref{table:obssummary}, the final baseline noise and the total integration times are listed for positions where \thcii\ is detected. Figures~\ref{figure:positions_N159} to \ref{figure:positions_N160} visualize the observed positions.

\section{Combined \thcii\ spectra and the \cii\ optical depth}

The left panels in Fig.~\ref{figure:spectra_selected} show the \twcii\ spectra (blue) for positions where the \thcii\ emission is detected (Table~\ref{table:obssummary}). For positions 1, 3, and 4, where the HFAV center pixel is observed at the same positions, we also show the obtained \oi\ 63\um\ spectra. Note that the beam size of \oi\ is smaller by a factor of 2.2 (6.3\arcsec). In the middle two columns, the \thcii\ F=1-0 and F=1-1 spectra are represented as red lines. The right panels show the combined \thcii\ spectra following \citet{Guevara2019}:

\begin{align}
T_\textrm{mb,13,tot}(\varv) &= \dfrac{\Sigma_{F,F^\prime}w_{\textrm{F}\rightarrow\textrm{F}^\prime}T_\textrm{mb,13}(\varv+\delta \varv_{\textrm{F}\rightarrow\textrm{F}^\prime})}{\Sigma_{F,F^\prime}w_{\textrm{F}\rightarrow\textrm{F}^\prime}s_{\textrm{F}\rightarrow\textrm{F}^\prime}}\\
w_{\textrm{F}\rightarrow\textrm{F}^\prime}&=\dfrac{s_{\textrm{F}\rightarrow\textrm{F}^\prime}}{\sigma_\textrm{rms}^2}
\end{align}

where $\delta \varv_{\textrm{F}\rightarrow\textrm{F}^\prime}$ is the velocity offset of the \thcii\ lines relative to the \twcii\ line, and $s_{\textrm{F}\rightarrow\textrm{F}^\prime}$ are the relative intensities of the hyperfine components ($\Sigma_{F,F^\prime}s_{\textrm{F}\rightarrow\textrm{F}^\prime}=1$ when using all three hyperfine components). We use $\delta \varv_{1\rightarrow 0}=-65.2$~\kms, $\delta \varv_{1\rightarrow 1}=63.2$~\kms, $s_{1\rightarrow 0}=0.25$, and $s_{1\rightarrow 1}=0.125$ \citep{Ossenkopf2013}. We composed the \thcii\ spectra only from the \thcii\ F=1-0 and F=1-1 spectra because F=2-1 is blended with the \twcii\ line for all sources. The equation scales up the sum of the detected hyperfine components to represent the full \thcii\ emission, taking into account that the F=2-1 line should contribute 62.5\%.

For optically thin \twcii\ emission, the expected spectrum for the combined \thcii\ is scaled as $T_\textrm{mb,13,tot}(\varv) = T_\textrm{mb,12}(\varv)/\alpha^+$ (blue lines in the right panels of Fig.~\ref{figure:spectra_selected}). We assume that the isotopic ratio of carbon ions $\alpha^+=$\twcplus/\thcplus\ equals the elemental abundance of \twc/\thc$=49$ \citep{Wang2009} for the LMC (see Sect.~\ref{subsec:isotopicratio}). The expected spectrum of individual \thcii\ hyperfine line is scaled by $s_{\textrm{F}\rightarrow\textrm{F}^\prime}$.

We derived the \twcii\ optical depth assuming that the excitation temperature of the \thcii\ and \twcii\ is the same \citep{Ossenkopf2013,Guevara2019}:

\begin{equation}
\frac{T_\textrm{mb,12}(\varv)}{T_\textrm{mb,13}(\varv)}=\frac{1-\exp(-\tau_{12}(\varv))}{1-\exp(-\tau_{13}(\varv))}
\end{equation}

where $\tau_{12}$ and $\tau_{13}$ are the optical depth of \twcii\ and \thcii, respectively, and $\tau_{12}=\alpha^+\tau_{13}$. We use the combined \thcii\ spectra and derived \twcii\ optical depth in each velocity bin. Figure~\ref{figure:tau12} shows the derived optical depth ($\tau_{12}$) with the errorbars.

\section{Discussion}

\subsection{Line profiles and optical depth}\label{subsec:opticaldepth}

At the N159~W \cii\ peak (position 1), N160 CO peak (position 4) and possibly N160~A (position 5), the \thcii\ spectra show an enhancement over the scaled \twcii\ spectra (Fig.~\ref{figure:spectra_selected}), which indicates that either the \twcii\ line is optically thick, or the isotopic ratio is lower than 49. As discussed in the following, the former is more likely because (1) the enhancement varies over different velocity bins, while it is reasonable to assume that different velocity components have the same isotopic ratio, (2) the \oi\ 63\um\ profile indicates self-absorption at the velocity where \thcii\ shows an enhancement (position 1 and 4), or (3) the peak velocity of the \thcii\ profile is consistent with the peak velocity of the \ci\ 492~GHz and \thco(3-2) lines (position 5).

At the N159~W \cii\ peak (position 1), there are two velocity components in the \twcii\ spectra (around $231$~\kms\ and $240$~\kms). For the stronger component around $231$~\kms, the \thcii\ intensity is larger than the scaled \twcii\ intensity, while the second velocity component around $240$~\kms\ does not indicate an enhanced \thcii\ emission; the peak intensity of the scaled \cii\ intensity for this velocity component is consistent with the noise level of \thcii. Since a variation of \twc/\thc\ within a physical scale of a few pc (14\arcsec\ of the beam size corresponds to 3.4~pc at the distance of the LMC; 50~kpc) has not been reported, we assume that the two velocity components have the same isotopic ratio, and attribute the difference in the \twcii/\thcii\ ratios between the two velocity components to the difference in their optical depths. The \oi\ 63\um\ emission indicates self-absorption around $231$~\kms, but none around $240$~\kms. This is consistent with \twcii\ being optically thick only for the $231$~\kms\ component. 

\begin{figure}
\centering
\includegraphics[width=\hsize]{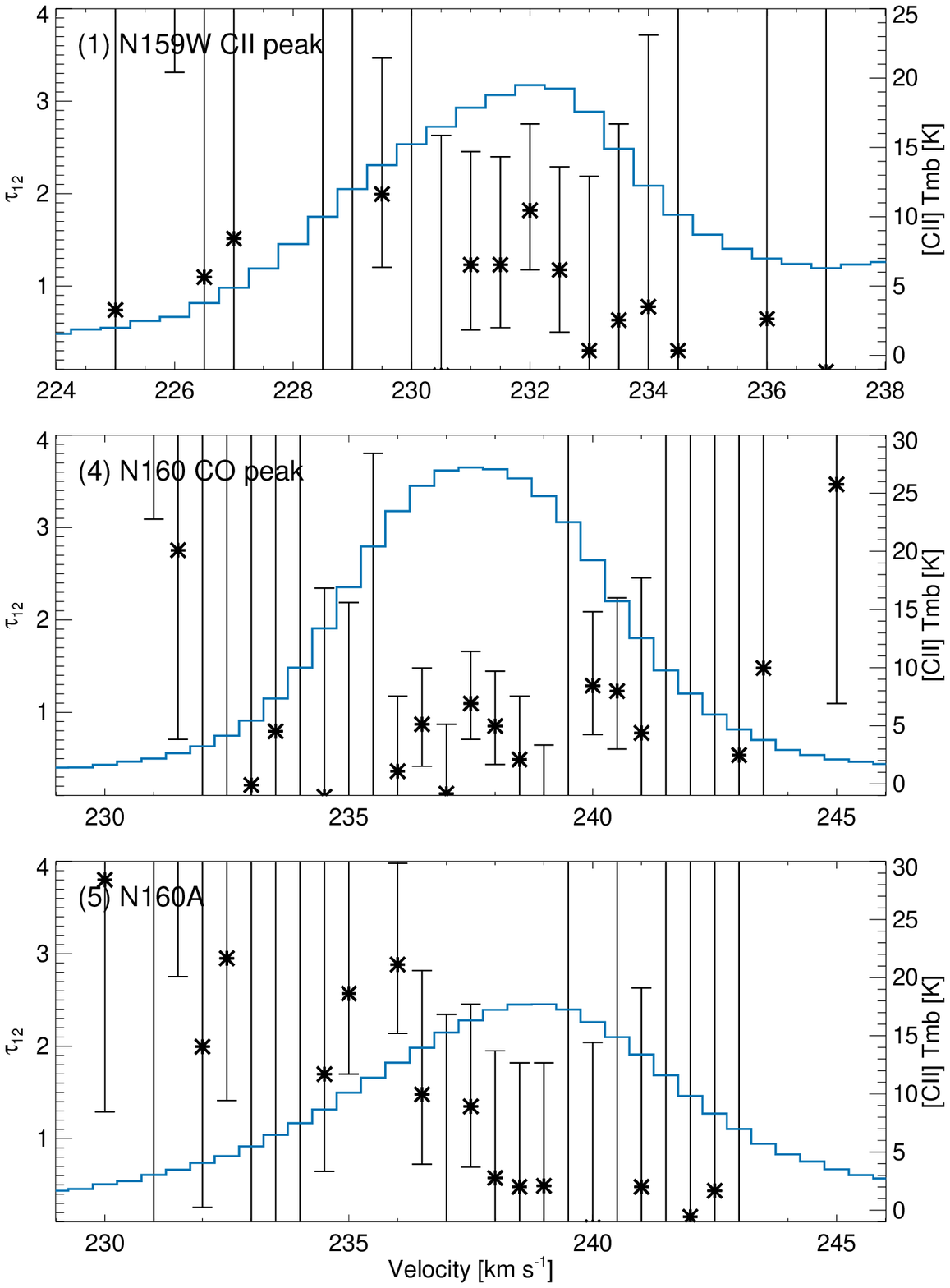}
\caption{Optical depth of the \cii\ emission ($\tau_{12}$) when assuming \twcplus/\thcplus$=49$ for each velocity bin in the three regions with enhanced \thcii. Asterisks indicate the derived $\tau_{12}$ together with the errorbars. Blue lines show the \cii\ emission profiles.}
\label{figure:tau12}
\end{figure}

The line ratio at the velocity bin of the \cii\ peak of position 1 translates into $\tau_{12}=1.8^{+0.9}_{-0.6}$ (Fig.~\ref{figure:tau12}), and a few velocity bins around the peak indicate an optical depth around unity. These values are similar to those of M43 and Horsehead measured by \citet{Guevara2019}, the mean value in the Orion molecular cloud \citep[1.3,][]{Goicoechea2015}, and somewhat lower than that of the Orion Bar \citep[$\sim3$,][]{Ossenkopf2013} and the bright shell confining Orion's Veil bubble \citep[3.5,][]{Pabst2019}.

The \thcii\ profile is narrower than the \twcii\ line for the velocity component around 231\,\kms. A single Gaussian fit gives a width of 3.6\,\kms\ for \thcii\ and 6.8\,\kms\ for \twcii. This difference is larger than expected from the optical depth broadening. For a Gaussian velocity distribution and a line-center optical depth of $\hat{\tau}<10$, the increase in the line width can be approximated as $1+0.115 \hat{\tau}$ \citep{Ossenkopf2013}. For $\hat{\tau}=1.8$, the optical depth broadens the line only by 20\%. The excess of the measured \twcii\ broadening compared to the optical depth broadening is consistent with the picture in \citet{Okada2019}; our beam includes several PDR components that are spatially separated and/or are in different physical phases, and each component contributes to a certain velocity range in the observed line profiles depending on their dynamics. Individual components have different \cii\ optical depths, and we see the enhancement of \thcii\ only for the components with a significant \cii\ optical depth. Thus \thcii\ is much narrower than \twcii. From the dataset presented in \citet{Okada2019}, we extracted the \ci\ 492~GHz, CO(4-3), and \thco(3-2) spectra at this position with 0.5\,\kms\ velocity resolution and 20\arcsec (for \thco(3-2)) or 16\arcsec\ (for two other lines) spatial resolution, and confirm that the line widths of these lines are also similar to that of \thcii\ (Fig.~\ref{figure:spectra_co_ci}).

At the N160 CO peak (position 4), the \thcii\ intensity is also higher than the scaled \twcii\ intensity, and the \oi\ 63\um\ profile indicates a large optical depth at the peak velocity of the \twcii\ line, although it is less clear than position 1. The derived $\tau_{12}$ at a few velocity bins around the peak is around unity. At N160~A (position 5), we do not see a clear enhancement of \thcii\ compared to the scaled \twcii\ at the peak velocity of \twcii, but one might be present around 235--237\,\kms. When fitting a single Gaussian, the central velocity is at 238.5\,\kms\ for \twcii\ and 237.4\,\kms\ for \thcii, and the fit to \ci\ 492 GHz and \thco(3-2) gives a closer central velocity (237.5\,\kms\ and 237.9\,\kms, respectively, see also Fig.~\ref{figure:spectra_co_ci}). This supports the suggestion that the \thcii\ enhancement around 235--237\,\kms\ is real. Fig.~\ref{figure:tau12} shows that the velocity bins around the \cii\ peak indicates an optically thin \cii\ emission, but $\tau_{12}$ of 1 to 4 is suggested at the blue wing (around 235\,\kms). 

At the other two positions (2 and 3), \thcii\ is marginally detected, and it is consistent with optically thin \cii\ emission within the noise level.

\subsection{Isotopic ratio} \label{subsec:isotopicratio}

The estimate of the optical depth of the \twcii\ line is based on an assumption of the isotopic ratio \twc/\thc, which is not as well studied in the LMC as in our Galaxy. \citet{Wang2009} obtain \twc/\thc$=49\pm 5$ in N113 in the LMC, which is presumably the most accurate carbon isotope ratio determined for the LMC because they use optically thin lines. It is consistent with previous measurements in other regions in the LMC; $50^{+25}_{-20}$ for N159 \citep{Johansson1994} and $35\pm 21$ for 30~Dor-10 \citep{Heikkila1999}. In the Galaxy, the \twc/\thc\ ratio increases along the Galactocentric distance \citep{WilsonRood1994}. The above value in the LMC is lower than the value for the local ISM in the solar neighborhood and close to the values in the inner Galaxy \citep{WilsonRood1994}, which is inconsistent with a pure metallicity dependence \citep{Wang2009}. \thc\ is a secondary nuclear fusion product, which is converted from \twc\ at the red giant stage \citep{Wilson1999}, and ejected into the ISM with a time delay. Therefore a low \twc/\thc\ can be explained by old stellar populations in the LMC \citep{Wang2009}.

\begin{figure}
\centering
\includegraphics[width=\hsize]{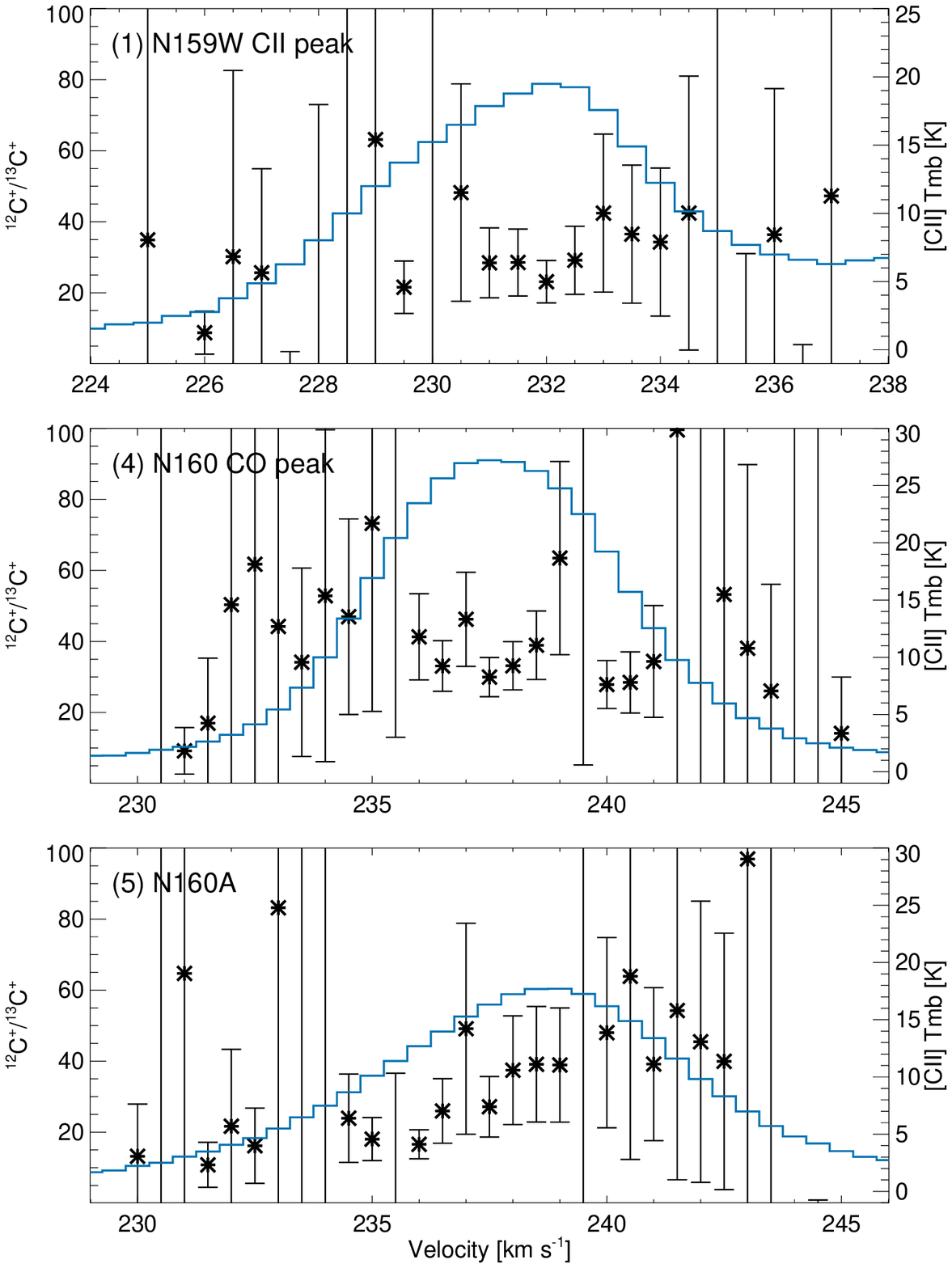}
\caption{Estimate of $\alpha^+=$\twcplus/\thcplus\ when assuming \twcii\ is optically thin. Asterisks indicate the derived $\alpha^+$ together with the errorbars. Blue lines show the \cii\ emission profiles.}
\label{figure:alpha}
\end{figure}

In Fig.~\ref{figure:alpha}, we estimated $\alpha^+=$\twcplus/\thcplus\ when assuming \twcii\ is optically thin for each velocity bin at the three positions discussed in Sect.~\ref{subsec:opticaldepth}. As discussed above, we assume that $\alpha^+$ is constant over different velocity components. At positions 1 (N159~W) and 5 (N160~A), $\alpha^+$ of 20--30 over the whole velocity range is not excluded when taking into account the noise level, but we do see a systematic trend across the velocity; a dip in the estimated $\alpha^+$ around the \cii\ peak at position 1 and a gradient at 235--240\,\kms\ at position 5. Also the estimated $\alpha^+$ of 20--30 is lower than the measured \twc/\thc\ in N159 \citep{Johansson1994}. Therefore the scenario of optically thick \twcii\ discussed in Sect.~\ref{subsec:opticaldepth} is more likely.

\subsection{Fractionation}

Because the fractionation reaction from \thcplus\ to \twcplus\ is an exothermal reaction \citep[the exothermicity is 35~K;][]{Langer1984}, \thcplus\ tends to be underabundant with respect to \twcplus\ at low temperatures, namely at higher $A_V$ in a PDR. \citet{RoelligOssenkopf2013} calculated the fractionation in the clump integrated intensity of \twcii\ and \thcii, and showed that only models with low UV fields ($\chi\leq100$) and high densities ($n\geq10^5$\,\cc) trace the chemical fractionation, which would result in $\alpha<\alpha^+$ and that the derived \twcii\ optical depth is a lower limit. On the other hand, it is unlikely that the fractionation is significant in the regions in this study because the estimated PDR properties are either low density and low UV field, or high density and high UV field \citep{Okada2019}.

\subsection{Metallicity effect}

Model predictions do not indicate a clear metallicity dependence of the \cii\ optical depth. For a high density PDR clump where the \cplus\ layer consists of a thin surface, the surface brightness of the \cii\ emission is almost constant with the metallicity \citep{Roellig2006}, because the surface brightness is proportional to the thickness of the layer and the density of \cplus\ ions, with the former being proportional to the inverse of the metallicity due to the dust extinction and the latter being proportional to the metallicity as a first order approximation. The derived optical depths in the LMC regions are similar to those of some Galactic regions \citep[M43 and Horsehead measured by][]{Guevara2019}. However this study does not provide enough statistics to conclude that we do not see a metallicity dependence in the \cii\ optical depth.

\subsection{Implications for the interpretation of the \cii\ intensity}

The impact of the optical depth effect depends strongly on the actual geometry of the sources. Most PDR models compute the \cii{} intensity emerging from the surface of an externally illuminated clump or a face-on plane-parallel structure. For this geometry the radiative transfer computation takes the optical depth correctly into account. However, when the \cii{} emission stems from a mixture of components including surfaces illuminated from the back (negative temperature gradient toward the observer) or an ensemble of clumps we can no longer simply sum up the intensity from all surfaces. Instead, we would have to run a full three-dimensional radiative transfer computation \citep[e.g.][]{Andree-Labsch2017}. It makes a significant difference when many clumps overlap along the line-of-sight even when the optical depth of each clump is only about unity. The interpretation of the optically thick \cii{} emission in terms of classical PDR model predictions will result in incorrect parameters.

To assess the impact of the optical depth on interpretations of the \cii\ emission in distant galaxies, we averaged over the available large maps in Orion \citep{Pabst2019} and M17 (Guevara et al. in prep.) to derive an averaged optical depth of \twcii. At the distance of LMC, the map sizes of about one degree (7.2~pc at Orion) and 2.5\arcmin (1.4~pc at M17) would correspond to 30\arcsec\ and 6\arcsec, respectively. The average optical depths are $\sim 1$ and $\sim 3$, respectively. These values are lower than the results from smaller areas \citep{Pabst2019,Guevara2019} but still moderately optically thick. To obtain representative values for other nearby galaxies we would need to average much larger areas. Such data are not available so far. However, our study indicates that the \cii\ emission in distant galaxies can have an optical depth of about unity, which leads to an underestimate of the \cii{} intensity by a factor two when assuming an optically thin scenario.

\section{Summary}
We detected \thcii\ F=1-0 and F=1-1 emissions in N159 and N160 in the LMC for the first time. Assuming an isotopic ratio \twcplus/\thcplus\ of 49, the optical depth of \twcii\ is estimated as 1--3 at the peak velocities. Although the possibility of an optically thin \twcii\ emission with a lower isotopic ratio is not quantitatively excluded, the fact that two velocity components in N159 have different intensity ratios of \twcii/\thcii, a narrower line profile of \thcii\ than \twcii, and a self-absorption of the \oi\ 63\um\ at the peak of the \thcii\ line favor an interpretation with an optically thick \twcii\ emission. This study indicates that the \cii\ intensity in distant galaxies can be underestimated by about a factor of 2.

\begin{acknowledgements}
This work is based on observations made with the NASA/DLR Stratospheric Observatory for Infrared Astronomy (SOFIA). SOFIA is jointly operated by the Universities Space Research Association, Inc. (USRA), under NASA contract NAS2-97001, and the Deutsches SOFIA Institut (DSI) under DLR contract 50 OK 0901 to the University of Stuttgart. This work is carried out within the Collaborative Research Centre 956, sub-project A4 and C1, funded by the Deutsche Forschungsgemeinschaft (DFG) – project ID 184018867.
\end{acknowledgements}

\begin{appendix}

\section{Additional figures}

\begin{figure}
\centering
\includegraphics[width=0.9\hsize]{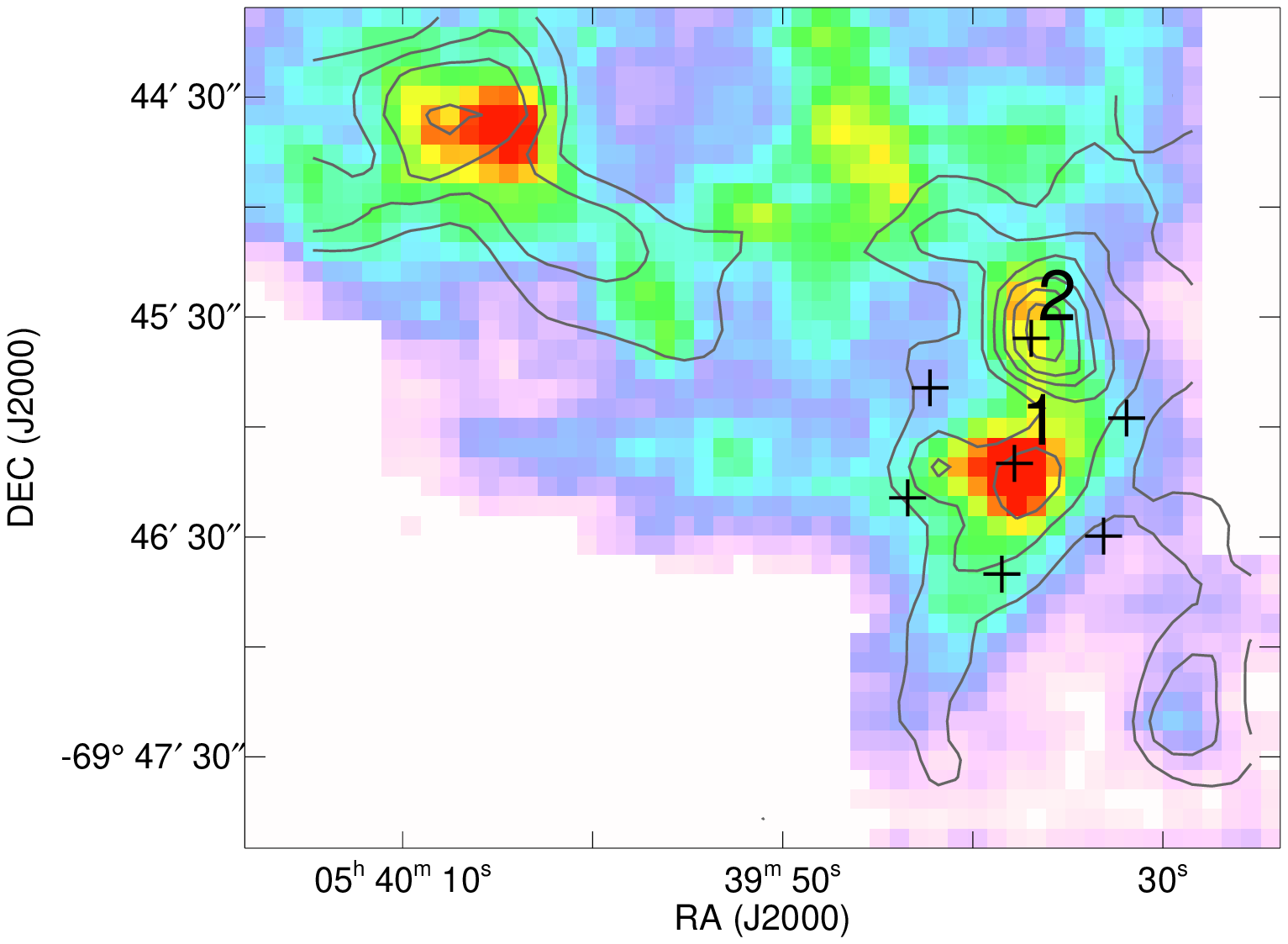}
\caption{Footprints of the LFA seven pixels in the \thcii\ observations in N159 overlaid with integrated intensities of \cii\ (color) and CO(4-3) (contours) \citep{Okada2019}. The labeled numbers correspond to the position IDs in Table~\ref{table:obssummary}.}
\label{figure:positions_N159}
\end{figure}

\begin{figure}
\centering
\includegraphics[width=0.9\hsize]{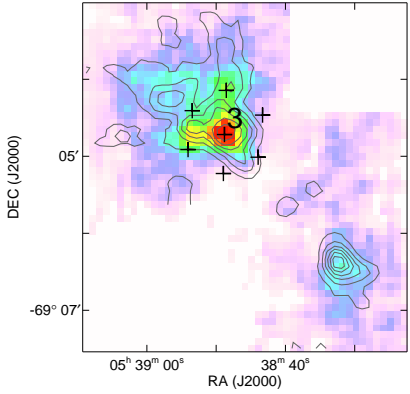}
\caption{Same as Fig.~\ref{figure:positions_30Dor} but for 30~Dor.}
\label{figure:positions_30Dor}
\end{figure}

\begin{figure}
\centering
\includegraphics[width=0.9\hsize]{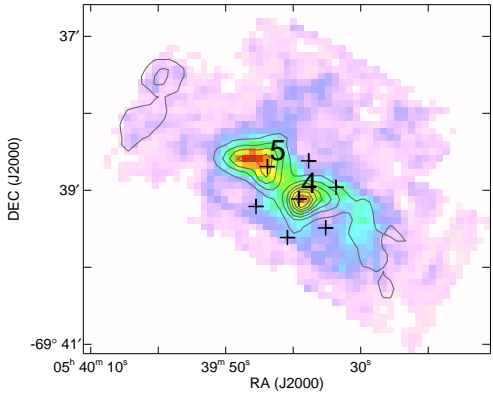}
\caption{Same as Fig.~\ref{figure:positions_N159} but for N160.}
\label{figure:positions_N160}
\end{figure}

Figures~\ref{figure:positions_N159} to \ref{figure:positions_N160} visualize the observed positions of the LFA seven pixels overlaid to the \cii\ and CO(4-3) maps in three regions. Five positions where \thcii\ is detected and discussed in this study are marked.

\begin{figure}
\centering
\includegraphics[width=0.9\hsize]{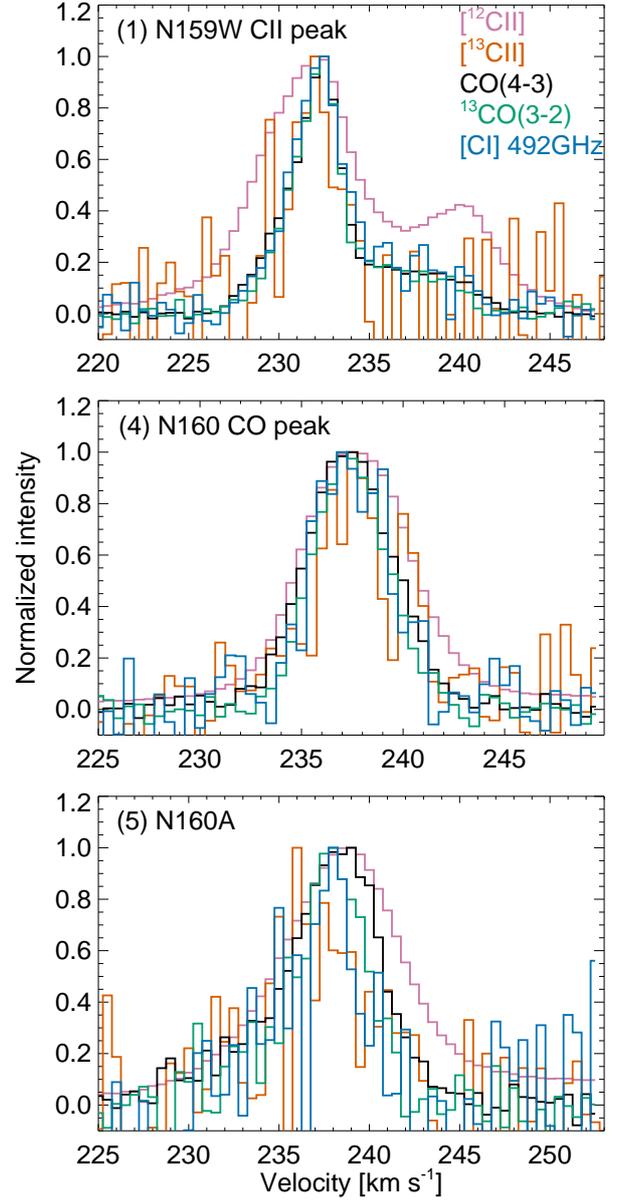}
\caption{Normalized spectra of CO(4-3), \thco(3-2), \ci\ 492~GHz \citep{Okada2019} together with the \twcii\ and \thcii\ emissions in this study at three positions with enhanced \thcii. \thco(3-2) has a spatial resolution of 20\arcsec\ and CO(4-3) and \ci\ 492~GHz have a spatial resolution of 16\arcsec.}
\label{figure:spectra_co_ci}
\end{figure}

Figure~\ref{figure:spectra_co_ci} show the normalized spectra of CO(4-3), \thco(3-2), \ci\ 492~GHz obtained from the dataset presented in \citet{Okada2019} together with the \twcii\ and \thcii\ emissions in this study. The \thco(3-2) spectra are extracted with the spatial resolution of 20\arcsec, CO(4-3) and \ci\ 492~GHz spectra are extracted with the spatial resolution of 16\arcsec\ at individual positions where the deep \thcii\ observations were executed.

\end{appendix}

\end{document}